\documentclass[manuscript]{biometrika}

\usepackage{graphicx}
\usepackage[utf8]{inputenc}
\usepackage{url}
\usepackage{mathrsfs}
\usepackage{amsmath,amsfonts,amssymb}
\usepackage[subnum]{cases}
\usepackage{times}

\def\Pr{{\rm pr}}
\def\E{{\rm \it E}}

\graphicspath{{./Figures/}} \title{Conditional simulation of max-stable processes}

\author{C. Dombry, F. \'Eyi-Minko,}

\affil{Laboratoire de Mathématiques et Application, Université de
  Poitiers, Téléport 2, BP 30179, F-86962 Futuroscope-Chasseneuil
  cedex, France \email{clement.dombry@math.univ-poitiers.fr},
  \email{frederic.eyi.minko@math.univ-poitiers.fr}}

\author{\and M. Ribatet}

\affil{Department of Mathematics, Université Montpellier 2, 4 place
  Eugène Bataillon, 34095 cedex 2 Montpellier, France
  \email{mathieu.ribatet@math.univ-montp2.fr}}

\def\footnote#1{{}}

\begin{document}
\maketitle

\begin{abstract}
  Since many environmental processes such as heat waves or
  precipitation are spatial in extent, it is likely that a single
  extreme event affects several locations and the areal modelling of
  extremes is therefore essential if the spatial dependence of
  extremes has to be appropriately taken into account. This paper
  proposes a framework for conditional simulations of max-stable
  processes and give closed forms for Brown--Resnick and Schlather
  processes. We test the method on simulated data and give an
  application to extreme rainfall around Zurich and extreme
  temperature in Switzerland. Results show that the proposed framework
  provides accurate conditional simulations and can handle real-sized
  problems.
\end{abstract}

\section{Introduction}
\label{sec:introduction}

Max-stable processes arise naturally when studying extremes of
stochastic processes and therefore play a major role in the
statistical modelling of spatial extremes
\citep{Buishand2008,Padoan2010,Davison2011}. Although a different
spectral characterization of max-stable processes exists
\citep{deHaan1984}, for our purposes the most useful representation is
\citep{Schlather2002}
\begin{equation}
  \label{eq:schlatherCharac}
  Z(x) = \max_{i \geq 1} \zeta_i Y_i(x), \qquad x \in \mathbb{R}^d,
\end{equation}
where $\{\zeta_i\}_{i \geq 1}$ are the points of a Poisson process on
$(0, \infty)$ with intensity $\mbox{d$\Lambda$}(\zeta) = \zeta^{-2}
\mbox{d$\zeta$}$ and $Y_i$ are independent replicates of a
non-negative continuous sample path stochastic process $Y$ such that
$\E\{Y(x)\} = 1$ for all $x \in \mathbb{R}^d$. It is well known that
$Z$ is a max-stable process on $\mathbb{R}^d$ with unit Fréchet
margins
\citep{deHaan2006b,Schlather2002}. Although~\eqref{eq:schlatherCharac}
takes the pointwise maximum over an infinite number of points
$\{\zeta_i\}$ and processes $Y_i$, it is possible to get approximate
realizations from $Z$ \citep{Schlather2002,Oesting2011}.

Based on~\eqref{eq:schlatherCharac} several parametric max-stable
models have been proposed
\citep{Schlather2002,Brown1977,Kabluchko2009,Davison2011} and share
the same finite dimensional distribution functions
\begin{equation*}
  % \label{eq:CDFMaxStab}
  \Pr\{Z(x_1) \leq z_1, \ldots, Z(x_k) \leq z_k\} = \exp\left[ -
    \E\left\{\max_{j=1,\ldots, k} \frac{Y(x_j)}{z_j} \right\} \right],
\end{equation*}
where $k \in \mathbb{N}$, $z_1, \ldots, z_k > 0$ and $x_1, \ldots,
x_k \in \mathbb{R}^d$.

Apart from the Smith model \citep{Genton2011}, only the bivariate
cumulative distribution functions are explicitly known. To bypass this
hurdle, \citet{deHaan2006} propose a semi-parametric estimator and
\citet{Padoan2010} suggest the use of the maximum pairwise likelihood
estimator.

Paralleling the use of the variogram in classical geostatistics, the
extremal coefficient function \citep{Schlather2003,Cooley2006}
\begin{equation*}
  % \label{eq:extCoeffFct}
  \theta(x_1 - x_2) = - z \log \Pr\{Z(x_1) \leq z, Z(x_2) \leq z\}
\end{equation*}
is widely used to summarize the spatial dependence of extremes. It
takes values in the interval $[1, 2]$; the lower bound indicates
complete dependence, and the upper one
independence.

The last decade has seen many advances to develop a geostatistic of
extremes and software is available to practitioners
\citep{maxLinear,RandomFields,SpatialExtremes}. However an important
tool currently missing is conditional simulation of max-stable
processes. In classical geostatistic based on Gaussian models,
conditional simulations are well established \citep{Chiles1999} and
provide a framework to assess the distribution of a Gaussian random
field given values observed at fixed locations. For example,
conditional simulations of Gaussian processes have been used to model
land topography \citep{Mandelbrot1982}.

Conditional simulation of max-stable processes is a long-standing
problem \citep{Davis1989,Davis1993}. \citet{Wang2011} provide a first
solution, but their framework is limited to processes having a
discrete spectral measure and thus may be too restrictive to
appropriately model the spatial dependence in complex situations.

Based on the recent developments on the regular conditional
distribution of max-infinitely divisible processes, the aim of this paper
is to provide a methodology to get conditional simulations of
max-stable processes with continuous spectral measures. More formally
for a study region $\mathcal{X} \subset \mathbb{R}^d$, our goal is to
derive an algorithm to sample from the regular conditional
distribution of $Z \mid \{Z(x_1) = z_1, \ldots, Z(x_k) = z_k\}$ for
some $z_1, \ldots, z_k > 0$ and $k$ conditioning locations $x_1,
\ldots, x_k \in \mathcal{X}$.

\section{Conditional simulation of max-stable processes}
\label{sec:cond-simul-brown}

\subsection{General framework}
\label{sec:general-framework}

This section reviews some key results of an unpublished paper
available from the first author with a particular emphasis on
max-stable processes. Our goal is to give a more practical
interpretation of their results from a simulation perspective. To this
aim, we recall two key results and propose a procedure to get conditional
realizations of max-stable processes.

Let $\mathbb{R}^\mathcal{X}$ be the space of  on $\mathcal{X}\subset\mathbb{R}^d$ and
let $\Phi =\{\varphi_i\}_{i\geq 1}$ be a Poisson point process on
$\mathbb{R}^\mathcal{X}$ where $\varphi_i(x) = \zeta_i Y_i(x)$ $(i = 1, 2,
\ldots)$ with $\zeta_i$ and $Y_i$ as in~\eqref{eq:schlatherCharac}.
We write $f({x}) = \{f(x_{1}), \ldots, f(x_k)\}$ for all random
functions $f\colon \mathcal{X} \to \mathbb{R}$ and ${x} = (x_1,
\ldots, x_k) \in \mathcal{X}^k$. It is not difficult to show that for
all Borel set $A \subset \mathbb{R}^k$,
the Poisson point process $\{\varphi_i({x})\}_{i \geq 1}$ defined on
$\mathbb{R}^k$ has intensity measure
\begin{equation*}
  \Lambda_x(A) = \int_0^{\infty} \Pr \left\{ \zeta Y(x) \in A \right\}
  \zeta^{-2} \mbox{d$\zeta$}.
\end{equation*}
The point process $\Phi$ is called regular if the intensity measure $\Lambda_x$ has an intensity
function $\lambda_x$, i.e., $\Lambda_x(\mbox{d$z$})=\lambda_x(z)\mbox{d$z$}$, for all $x\in \mathcal{X}^k$.

The first key point is that provided  the point process $\Phi$ is regular, the intensity function
$\lambda_x$ and the conditional intensity function 
\begin{equation}
  \label{eq:condIntensityFunction}
  \lambda_{s \mid x, z}(u) = \frac{\lambda_{(s,x)}(u,
    z)}{\lambda_x(z)}, \qquad (s, x) \in \mathcal{X}^{m+k}, \qquad u \in \mathbb{R}^{m},\qquad z\in (0,+\infty)^k,
\end{equation}
drives how the conditioning terms $\{Z(x_j) = z_j\}$ $(j=1, \ldots,k)$ are met.

The second key point is that, conditionally on $Z(x) = z$, the Poisson
point process $\Phi$ can be decomposed into two independent 
point processes, say $\Phi = \Phi^- \cup \Phi^+$, where
\begin{align*}
  \Phi^- &= \{\varphi \in \Phi \colon \text{$\varphi(x_i) < z_i$
    for all $i = 1, \ldots, k$}\},\\
  \Phi^+ &= \{\varphi \in \Phi \colon \text{$\varphi(x_i) = z_i$
    for some $i = 1,\ldots,k$}\}.
\end{align*}

Before introducing a procedure to get conditional realizations of
max-stable processes, we introduce notation and make connections with
the pioneering work of \citet{Wang2011}.

A function $\varphi \in \Phi^+$ such that $\varphi(x_i) = z_i$ for
some $i \in \{1, \ldots, k\}$ is called an extremal function
associated to $x_i$ and denoted by $\varphi_{x_i}^+$. It is easy to show that
there exists almost surely a unique extremal function associated to $x_i$. 
Although $\Phi^+ = \{\varphi_{x_1}^+,\ldots, \varphi_{x_k}^+\}$ almost surely, 
it might happen that a single extremal function contributes to the random vector
$Z({x})$ at several locations $x_i$, e.g., $\varphi^+_{x_1} =
\varphi^+_{x_2}$. To take such repetitions into account, we define a
random partition $\theta = (\theta_1, \ldots, \theta_\ell)$ of the set
$\{x_1, \ldots, x_k\}$ into $\ell=|\theta|$ blocks and extremal
functions $(\varphi_1^+,\ldots,\varphi_\ell^+)$ such that $\Phi^+ =
\{\varphi_1^+, \ldots, \varphi_\ell^+\}$ and $\varphi_{j}^{+} (x_i) =
z_i$ if $x_i \in \theta_j$ and $\varphi_{j}^{+} (x_i) < z_i$ if $x_i
\notin \theta_j$ $(i=1, \ldots, k;\ j=1, \ldots,
\ell)$. \citet{Wang2011} call the partition $\theta$ the hitting
scenario.  The set of all possible partitions of $\{x_1, \ldots,
x_k\}$, denoted $\mathscr{P}_k$, identifies all possible hitting
scenarios.

From a simulation perspective, the fact that $\Phi^-$ and $\Phi^+$ are
independent given $Z(x) = z$ is especially convenient and suggests a
three--step procedure to sample from the conditional distribution of
$Z$ given $Z(x) = z$.

\begin{theorem}
  \label{thm:3stepsCondSim}
  Suppose that the point process $\Phi$ is regular and let $(x,s) \in \mathcal{X}^{k+m}$. For $\tau =
  (\tau_1, \ldots, \tau_\ell) \in \mathscr{P}_k$ and
  $j=1,\ldots,\ell$, define $I_j = \{i\colon x_i \in \tau_j\}$,
  ${x}_{\tau_j} = (x_i)_{i \in I_j}$, ${z}_{\tau_j} = (z_i)_{i \in
    I_j}$, ${x}_{\tau_j^c} = (x_i)_{i \notin I_j}$ and ${z}_{\tau_j^c}
  = (z_i)_{i \notin I_j}$. Consider the three--step procedure:
  \begin{step}
    Draw a random partition $\theta\in\mathscr{P}_k$ with distribution
    \begin{equation*}
      \pi_{x}({z}, \tau) = \Pr\left\{ \theta=\tau \mid
        Z({x}) = {z}\right\} =\frac{1}{C({x},
        {z})} \prod_{j=1}^{|\tau|}
      \lambda_{{x}_{\tau_j}}({z}_{\tau_j})
      \int_{\{{u}_j < {z}_{\tau_j^c}\}}
      \lambda_{{x}_{\tau_j^c} \mid {x}_{\tau_j},
        {z}_{\tau_j}}({u}_j) \mbox{d${u}_j$},
    \end{equation*}
    where the normalization constant is
    \begin{equation*}
      C({x}, {z}) = \sum_{\tilde\tau \in \mathscr{P}_k}
      \prod_{j=1}^{|\tilde\tau|}
      \lambda_{{x}_{\tilde\tau_j}}({z}_{\theta_j})
      \int_{\{{u}_j < {z}_{\tilde\tau_j^c}\}}
      \lambda_{{x}_{\tilde\tau_j^c } \mid {x}_{\tilde\tau_j},
        {z}_{\tilde\tau_j}}({u}_j) \mbox{d${u}_j$}.
    \end{equation*}
  \end{step}
  \begin{step}
    Given $\tau = (\tau_1, \ldots, \tau_\ell)$, draw $\ell$
    independent random vectors $\varphi_{1}^{+}({s}), \ldots,
    \varphi_{\ell }^{+}({s})$ from the distribution
    \begin{equation*}
      \Pr\left\{ \varphi_{j}^{+}({s}) \in
        \mbox{d${v}$} \mid Z({x}) =
        {z}, \theta=\tau \right\} = \frac{1}{C_j} \left\{ \int
        1_{\{{u} < {z}_{\tau_j^c}\}} \lambda_{({s},
          {x}_{\tau_j^c}) \mid {x}_{\tau_j},
          {z}_{\tau_j}} ({v}, {u})
        \mbox{d${u}$} \right\}\,\mbox{d${v}$}
    \end{equation*}
    where $1_{\{\cdot\}}$ is the indicator function and
    \begin{equation*}
      C_j = \int 1_{\{{u} < {z}_{\tau_j^c}\}}
      \lambda_{({s}, {x}_{\tau_j^c}) \mid {x}_{\tau_j},
        {z}_{\tau_j}} ({v}, {u})
      \mbox{d${u}$d${v}$},
    \end{equation*}
    and define the random vector $Z^+({s}) = \max_{j=1, \ldots,
      \ell} \varphi_{j}^{+}({s})$.
  \end{step}
  \begin{step}
    Independently draw a Poisson point process $\{\zeta_i\}_{i \geq
      1}$ on $(0, \infty)$ with intensity $\zeta^{-2} \mbox{d$\zeta$}$
    and $\{Y_i\}_{i\geq 1}$ independent copies of $Y$, and define the
    random vector
    \begin{equation*}
      Z^-({s}) = \max_{i \geq 1} \zeta_i Y_i(s) 1_{\{\zeta_i Y_i(x) < z \}}.
    \end{equation*}
  \end{step}
  Then the random vector $\tilde Z({s}) = \max \left\{
    Z^+({s}), Z^-({s}) \right\}$ follows the conditional
  distribution of $Z({s})$ given $Z({x}) = {z}$.
\end{theorem}

The corresponding conditional cumulative distribution function is
\begin{equation}
  \label{eq:condCDF}
  \Pr\left\{ Z({s}) \leq {a} \mid Z({x}) =
    {z} \right\}  = \frac{\Pr\{Z({s}) \leq {a},
    Z({x}) \leq {z}\}}{\Pr\{Z({x}) \leq
    {z}\}} \sum_{\tau \in \mathscr{P}_k}
  \pi_{x}({z}, \tau) \prod_{j=1}^{|\tau|}
  F_{\tau,j}({a}),
\end{equation}
where
\begin{equation*}
  F_{\tau,j}({a}) =   \Pr\left\{ \varphi_{j}^{+}({s})\leq {a} \mid Z({x}) =
    {z},\,\theta=\tau \right\}= \frac{\int_{\{{u} <
      {z}_{\tau_j^c}, {v} < {a}\}}
    \lambda_{ ({s}, {x}_{ \tau_j^c}) \mid  {x}_{\tau_j},
      {z}_{\tau_j}} ({v}, {u})
    \mbox{d${u}$d${v}$}}{ \int_{\{{u} <
      {z}_{\tau_j^c}\}} \lambda_{ {x}_{ \tau_j^c} |
      {x}_{\tau_j}, {z}_{\tau_j}} ({u})
    \mbox{d${u}$}}.
\end{equation*}
It is clear from~\eqref{eq:condCDF} that the conditional random field
$Z \mid \{Z(x) = z\}$ is not max-stable.

\subsection{Distribution of the extremal functions}
\label{sec:distr-extr-funct}

In this section we derive closed forms for the intensity function
$\lambda_x(z)$ and the conditional intensity function $\lambda_{s \mid
  x, z}(u)$ for two widely used max-stable processes; the
Brown--Resnick \citep{Brown1977,Kabluchko2009} and the Schlather
\citep{Schlather2002} processes. Details of the derivations of these
closed forms are given in the Appendix.

The Brown--Resnick process corresponds to the case where $Y(x) = \exp
\{W(x) - \gamma(x)\}$, $x \in \mathbb{R}^d$,
in~\eqref{eq:schlatherCharac} with $W$ a centered Gaussian process
with stationary increments, semi variogram $\gamma$ and such that
$W(o) = 0$ almost surely. For $x \in \mathcal{X}^k$ and provided the
covariance matrix $\Sigma_x$ of the random vector $W(x)$ is positive
definite, the intensity function is
\begin{align*}
  \lambda_x(z) &= C_x \exp\left(-\frac{1}{2} \log {z}^T Q_x \log 
    z + L_x \log z \right) \prod_{i=1}^k z_i^{-1}, \qquad z \in (0,
  \infty)^k,
\end{align*}
with $1_k = (1)_{i=1, \ldots, k}$, $\sigma^2_x = \{\sigma^2(x_i)\}_{i
  = 1, \ldots, k}$,
\begin{align*}
  Q_x &= \Sigma_x^{-1} - \frac{\Sigma_x^{-1} 1_k 1_k^T
    \Sigma_x^{-1}}{1_k^T \Sigma_x^{-1} 1_k}, \qquad L_x = \frac{1}{2} 
  \left(\frac{1_k^T \Sigma_x^{-1} \sigma^2_x - 2}{1_k^T \Sigma_x^{-1} 
      1_k} 1_k^T - {\sigma^2_x}^T \right) \Sigma_x^{-1},\\
  C_x &= (2\pi)^{(1-k)/2} |\Sigma_x|^{-1/2} (1_k^T \Sigma_x^{-1}
  1_k)^{-1/2} \exp\left\{\frac{1}{2} \frac{(1_k^T \Sigma_x^{-1}
      \sigma^2_x - 1)^2}{1_k^T \Sigma_x^{-1} 1_k} -
    \frac{1}{2} {\sigma^2_x}^T \Sigma_x^{-1} \sigma^2_x \right\},
\end{align*}
and for all $(s, x) \in \mathcal{X}^{m+k}$, $(u, z) \in
(0,\infty)^{m+k}$ and provided the covariance matrix
$\Sigma_{(s,x)}$ is positive definite, the conditional intensity
function corresponds to a multivariate log-normal probability
density function
\begin{equation*}
  \lambda_{s \mid x, z}({u}) = (2\pi)^{-m/2} |\Sigma_{s \mid x}|^{-1/2}
  \exp\left\{-\frac{1}{2} (\log u - \mu_{s \mid x, z})^T \Sigma_{s
      \mid x}^{-1} (\log u - \mu_{s \mid x, z}) \right\} \prod_{i=1}^m
  u_i^{-1},
\end{equation*}
where $\mu_{s \mid x, z}\in \mathbb{R}^m$ and $\Sigma_{s \mid x}$
are the mean and covariance matrix of the underlying normal
distribution and are given by
\begin{align*}
  \Sigma_{s \mid x}^{-1} = J_{m,k}^{T} Q_{(s, x)} J_{m,k}, \qquad 
  \mu_{s \mid x, z} = \left\{ L_{(s, x)} J_{m,k} - \log z^T
    \tilde{J}_{m,k}^{\,T} Q_{(s, x)} J_{m,k} \right\} \Sigma_{s \mid
    x},
\end{align*}
with
\begin{equation*}
  J_{m,k} =
  \begin{bmatrix}
    \mbox{Id}_m\\
    {0}_{k,m}
  \end{bmatrix},
  \qquad  
  \tilde J_{m,k}=
  \begin{bmatrix}
    0_{m,k}\\
    \mbox{Id}_k
  \end{bmatrix},
\end{equation*}
where $\mbox{Id}_k$ denotes the $k \times k$ identity matrix and
$0_{m,k}$ the $m \times n$ null matrix.

The Schlather process considers the case where $Y(x) = (2 \pi)^{1/2}
\max\{0, \varepsilon(x)\}$, $x \in \mathbb{R}^d$,
in~\eqref{eq:schlatherCharac} with $\varepsilon$ a standard Gaussian
process with correlation function $\rho$. The associated point process
$\Phi$ is not regular and it is more convenient to consider the
equivalent representation where $Y(x) = (2 \pi)^{1/2} \varepsilon(x)$,
$x \in \mathbb{R}^d$. For $x \in \mathcal{X}^k$ and provided the
covariance matrix $\Sigma_x$ of the random vector $\varepsilon(x)$ is
positive definite, the intensity function is
\begin{equation*}
  \label{eq:densityPPP}
  \lambda_x(z) = \pi^{-(k - 1)/2} |\Sigma_{x}|^{-1/2}
  a_x(z)^{-(k+1)/2} \Gamma\left(\frac{k+1}{2} \right), \qquad z \in
  \mathbb{R}^k,
\end{equation*}
where $a_x(z) = z^T \Sigma_x^{-1} z$.

For $(s, x) \in \mathcal{X}^{m+k}$, $(u, z) \in \mathbb{R}^{m+k}$ and
provided that the covariance matrix $\Sigma_{(s, x)}$ is positive
definite, the conditional intensity function $\lambda_{s\mid x,
  z}(u)$ corresponds to the density of a multivariate Student
distribution with $k+1$ degrees of freedom, location parameter $\mu =
\Sigma_{s:x} \Sigma_x^{-1} z$, and scale matrix
\begin{equation*}
  \tilde \Sigma = \frac{a_x(z)}{k+1} \left(\Sigma_s - \Sigma_{s:x} 
    \Sigma_x^{-1} \Sigma_{x:s} \right), \qquad \Sigma_{(s, x)} =
  \begin{bmatrix}
    \Sigma_s & \Sigma_{s:x}\\
    \Sigma_{x:s} & \Sigma_x
  \end{bmatrix}.
\end{equation*}

\section{Markov chain Monte Carlo sampler}
\label{sec:markov-chain-monte}

The previous section introduced a procedure to get realizations from
the regular conditional distribution of max-stable processes. This
sampling scheme amounts to sample from a discrete distribution whose
state space corresponds to all possible partitions of the set of
conditioning points, see Theorem~\ref{thm:3stepsCondSim} step
1. Hence, even for a moderate number $k$ of conditioning locations,
the state space $\mathscr{P}_k$ becomes very large and the
distribution $\pi_x(z, \cdot)$ cannot be computed. It turns out that a
Gibbs sampler is especially convenient.

For $\tau \in \mathscr{P}_k$, let $\tau_{-j}$ be the restriction of
$\tau$ to the set $\{x_1, \ldots, x_k\} \setminus \{x_j\}$. Our goal
is to simulate from the conditional distribution
\begin{equation}
  \label{eq:fullCondDist}
  \Pr(\theta \in \cdot \mid \theta_{-j} = \tau_{-j}),
\end{equation}
where $\theta \in \mathscr{P}_k$ is a random partition which follows
the target distribution $\pi_x(z, \cdot)$.

Since the number of possible updates is always less than $k$, a
combinatorial explosion is avoided. Indeed for $\tau \in
\mathscr{P}_k$ of size $\ell$, the number of partitions $\tau^* \in
\mathscr{P}_k$ such that $\tau^*_{-j} = \tau_{-j}$ for some $j \in
\{1, \ldots, k\}$ is
\begin{equation*}
  b^+ = 
  \begin{cases}
    \ell &\text{if $\{x_j\}$ is a partitioning set of $\tau$},\\
    \ell + 1 & \text{if $\{x_j\}$ is not a partitioning set of
      $\tau$},
  \end{cases}
\end{equation*}
since the point $x_j$ may be reallocated to any partitioning set of
$\tau_{-j}$ or to a new one.

To illustrate consider the set $\{x_1, x_2, x_3\}$ and let $\tau =
(\{x_1, x_2\}, \{x_3\})$. Then the possible partitions $\tau^*$ such
that $\tau^*_{-2} = \tau_{-2}$ are $(\{x_1, x_2\}, \{x_3\})$,
$(\{x_1\}, \{x_2\}, \{x_3\})$, $(\{x_1\}, \{x_2, x_3\})$, while there
exists only two partitions such that $\tau^*_{-3} = \tau_{-3}$, i.e.,
$(\{x_1, x_2\}, \{x_3\})$, $(\{x_1, x_2, x_3\})$.

The distribution~\eqref{eq:fullCondDist} has nice properties. Since
for all $\tau^* \in \mathscr{P}_k$ such that $\tau^*_{-j} = \tau_{-j}$
we have
\begin{equation}
  \label{eq:condDistPart}
  \Pr[\theta = \tau^* \mid \theta_{-j} = \tau_{-j}] =
  \frac{\pi_{x}({z}, \tau^*)}{{\displaystyle
      \sum_{\tilde \tau \in \mathscr{P}_k} \pi_{x}({z},
      \tilde \tau)} 1_{\{\tilde \tau_{-j} = \tau_{-j}\}}} \propto
  \frac{\prod_{j=1}^{|\tau^*|} w_{\tau^*\!, j}}{\prod_{j=1}^{|\tau|}
    w_{\tau, j}},
\end{equation}
where
\begin{equation*}
  w_{\tau, j} = \lambda_{{x}_{\tau_j}}({z}_{\tau_j})
  \int_{\{{u} < {z}_{\tau_j^c}\}}
  \lambda_{{x}_{\tau^c_j} \mid {x}_{\tau_j},
    {z}_{\tau_j}}({u}) \mbox{d${u}$}.
\end{equation*}
Since many factors cancel out on the right hand side
of~\eqref{eq:condDistPart}, the Gibbs sampler is especially
convenient.

The most computationally demanding part of~\eqref{eq:condDistPart}
is the evaluation of the integral
\begin{equation*}
  \int_{\{{u} < {z}_{\tau_j^c}\}}
  \lambda_{{x}_{\tau_j^c} \mid {x}_{\tau_j},
    {z}_{\tau_j}}({u}) \mbox{d${u}$}.
\end{equation*}
For the Brown--Resnick and Schlather processes, we follow the lines of
\citet{Genz1992} and compute these probabilities using a separation of
variables method which provides a transformation of the original
integration problem to the unit hyper-cube. Further a quasi Monte
Carlo scheme and antithetic variable sampling is used to improve
efficiency.

Since it is not obvious how to implement a Gibbs sampler whose target
distribution has support $\mathscr{P}_k$, the remainder of this
section gives practical details. For any fixed locations $x_1, \ldots,
x_k \in \mathcal{X}$, we first describe how each partition of $\{x_1,
\ldots, x_k\}$ is stored.  To illustrate consider the set $\{x_1, x_2,
x_3\}$ and the partition $(\{x_1, x_2\}; \{x_3\})$. This partition is
defined as $(1, 1, 2)$, indicating that $x_1$ and $x_2$ belong to the
same partitioning set labeled $1$ and $x_3$ belongs to the
partitioning set $2$. There exist several equivalent notations for
this partition: for example one can use $(2, 2, 1)$ or $(1, 1,
3)$. Since there is a one-one mapping between $\mathscr{P}_k$ and the
set
\begin{equation*}
  \mathscr{P}_k^* = \left\{(a_1, \ldots, a_k)\colon i \in \{2, \ldots,
    k\}, \, 1=a_1 \leq a_i \leq \max_{1 \leq j < i} a_j + 1, \, a_i
    \in \mathbb{Z} \right\},
\end{equation*}
we shall restrict our attention to the partitions that live in
$\mathscr{P}_k^*$ and going back to our example we see that $(1, 1,
2)$ is valid but $(2, 2, 1)$ and $(1, 1, 3)$ are not.

For $\tau \in \mathscr{P}_k^*$ of size $\ell$, let $r_1 = \sum_{i=1}^k
1_{\{\tau_i = a_j\}}$ and $r_2 = \sum_{i=1}^k 1_{\{\tau_i = b\}}$,
i.e., the number of conditioning locations that belong to the
partitioning sets $a_j$ and $b$ where $b \in \{1, \ldots, b^+\}$ with
\begin{equation*}
  b^+ =
  \begin{cases}
    \ell &(r_1 = 1),\\
    \ell + 1 &(r_1 \neq 1).
  \end{cases}
\end{equation*}
Then the conditional probability distribution~\eqref{eq:condDistPart}
satisfies
\begin{numcases}{\Pr(\tau_j = b \mid \tau_i = a_i, \, i \neq j)
    \propto}
  \label{eq:condDist1}
  1 & $(b = a_j)$,\\
  \label{eq:condDist2}
  w_{\tau^*\!, b} / (w_{\tau, b} w_{\tau, a_j})
  & $(r_1 = 1, r_2 \neq 0, b \neq a_j)$,\\
  \label{eq:condDist3}
  w_{\tau^*\!, b} w_{\tau^*\!, a_j} / (w_{\tau, b} w_{\tau, a_j})
  & $(r_1 \neq 1, r_2 \neq 0, b \neq a_j)$,\\
  \label{eq:condDist4}
  w_{\tau^*\!, b} w_{\tau^*\!, a_j} / w_{\tau, a_j} & $(r_1 \neq 1, r_2 =
  0, b \neq a_j)$,
\end{numcases}
where $\tau^* = (a_1, \ldots, a_{j-1}, b, a_{j+1}, \ldots,
a_k)$. Although $\tau^*$ may not belong to $\mathscr{P}_k^*$, it
corresponds to a unique partition of $\mathscr{P}_k$ and we can use
the bijection $\mathscr{P}_k\to \mathscr{P}_k^*$ to recode $\tau^*$
into an element of
$\mathscr{P}_k^*$. In~\eqref{eq:condDist1}--\eqref{eq:condDist4} the
event $\{r_1 =1, r_2 = 0, b \neq a_j\}$ is missing since $\{r_1 = 1,
r_2 = 0\}$ implies that $\tau^* = \tau$, where the equality has to be
understood in terms of elements of $\mathscr{P}_k$, and this case has
been already taken into account with~\eqref{eq:condDist1}.

Once these conditional weights have been computed, the Gibbs sampler
proceeds by updating each element of $\tau$ successively. We use a
random scan implementation of the Gibbs sampler \citep{Liu1995}. More
precisely, one iteration of the random scan Gibbs sampler selects an
element of $\tau$ at random according to a given distribution, say
${p} = (p_1, \ldots, p_k)$, and then updates this element. Throughout
this paper we will use the uniform random scan Gibbs sampler for which
the selection distribution is assumed to be a discrete uniform
distribution, i.e., ${p} = (k^{-1}, \ldots, k^{-1})$.

\section{Simulation Study}
\label{sec:simulation-study}

\begin{table}
  \tbl{Sample path properties of the max-stable models. For the
    Brown--Resnick model, the variogram
    parameters are set to ensure that the extremal coefficient
    function satisfies $\theta(115) = \text{1$\cdot$7}$ while
    the correlation function parameters are set to ensure that
    $\theta(100) = \text{1$\cdot$5}$ for the Schlather model.}{%
    \begin{tabular}{lccccccc}
      & \multicolumn{3}{c}{Brown--Resnick: $\gamma(h) = (h /
        \lambda)^\kappa$} & & \multicolumn{3}{c}{Schlather: $\rho(h) =
        \exp\{-(h / \lambda)^\kappa\}$}\\
      & $\gamma_1$: Very wiggly & $\gamma_2$: Wiggly & $\gamma_3$:
      Smooth & & $\rho_1$: Very wiggly & $\rho_2$: Wiggly & $\rho_3$:
      Smooth\\
      $\lambda$ &  25 & 54 & 69 & & 208 & 144 & 128\\
      $\kappa$ & 0$\cdot$5 & 1$\cdot$0 & 1$\cdot$5 & & 0$\cdot$5 &
      1$\cdot$0 & 1$\cdot$5
    \end{tabular}}\label{tab:spatDepStruc}
\end{table}

In this section we check if our algorithm is able to produce realistic
conditional simulations of Brown--Resnick and Schlather processes. For
each model, we consider three different sample path properties, as
summarized in Table~\ref{tab:spatDepStruc}. These configurations were
chosen such that the spatial dependence structures are similar to our
applications in Section~\ref{sec:application}.

\begin{figure}
  \centering
  \includegraphics[width=\textwidth]{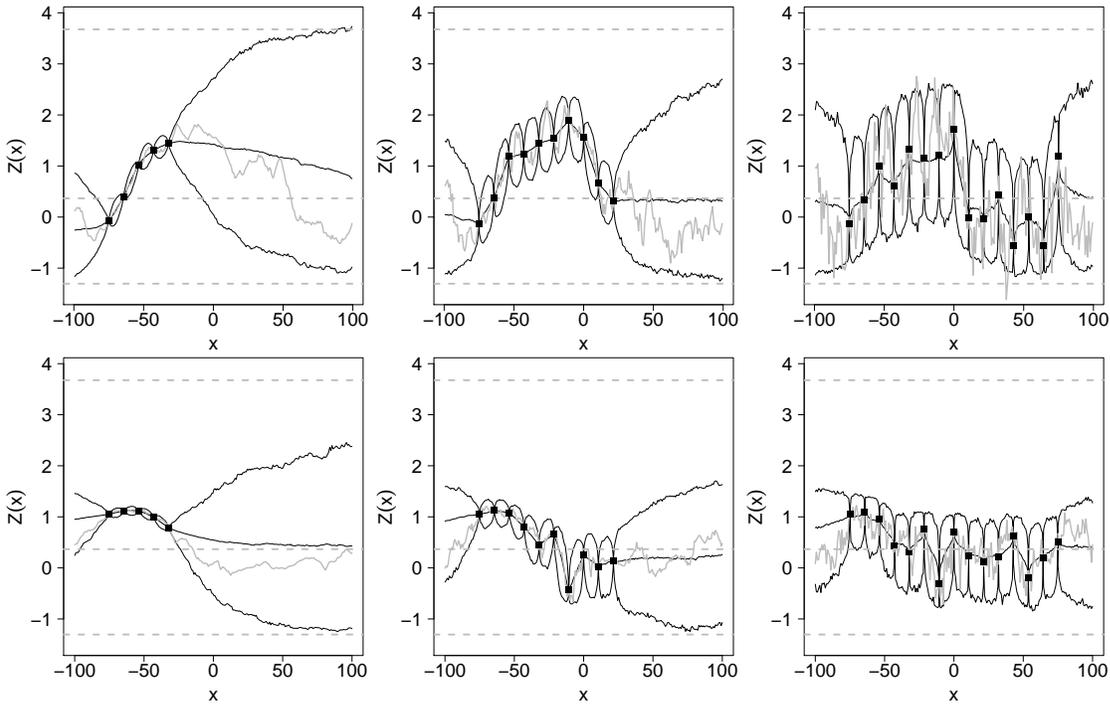}
  \caption{Pointwise sample quantiles estimated from 1000 conditional
    simulations of max-stable processes with standard Gumbel margins
    with $k = 5, 10, 15$ conditioning locations. The top row shows
    results for the Brown--Resnick models with semi variograms
    $\gamma_3, \gamma_2, \gamma_1$, from left to right. The bottom row
    shows results for the Schlather models with correlation functions
    $\rho_3, \rho_2, \rho_1$, from left to right. The solid black
    lines shows the pointwise 0$\cdot$025, 0$\cdot$5, 0$\cdot$975
    sample quantiles and the dashed grey lines that of a standard
    Gumbel distribution. The squares show the conditional points
    $\{(x_i, z_i)\}_{i=1, \ldots, k}$. The solid grey lines correspond
    the simulated paths used to get the conditioning events.}
  \label{fig:condSim5condLoc2}
\end{figure}

In order to check if our sampling procedure is accurate and given a
single conditional event $\{Z(x) =z\}$ for each configuration, we
generated $1000$ conditional realizations with standard Gumbel
margins. Figure~\ref{fig:condSim5condLoc2} shows the pointwise sample
quantiles obtained from these 1000 simulated paths and compares them
to unit Gumbel quantiles. As expected the conditional sample paths
inherit the regularity driven by the shape parameter $\kappa$ and there
is less variability in regions close to conditioning locations. Since
the considered Brown--Resnick processes are ergodic
\citep{Kabluchko2010}, for regions far away from any conditioning
location the sample quantiles converges to that of a standard Gumbel
distribution indicating that the conditional event has no
influence. This is not the case for the non-ergodic Schlather
processes. Most of the time the sample paths used to get the
conditional events belong to the 95\% pointwise confidence intervals,
corroborating that our sampling procedure seems to be accurate.

\begin{table}
  \tbl{Computational timings for conditional simulations of max-stable
    processes on a $50 \times 50$ grid defined on the square $[0,100
    \times 2^{1/2}]^2$ for a varying number $k$ of conditioning
    locations uniformly distributed over the region. The timings, in
    seconds, are mean values over 100 conditional simulations;
    standard deviations are reported in brackets.}{%
    \begin{tabular}{lccccccccc}
      & \multicolumn{4}{c}{Brown--Resnick: $\gamma(h) = (h / 25)^{0\cdot5}$} & &
      \multicolumn{4}{c}{Schlather: $\rho(h) = \exp\left\{-(h /
          208)^{0\cdot50}\right\}$}\\
      & Step 1 & Step 2 & Step 3 & Overall & & Step 1 & Step 2 & Step 3
      & Overall\\
      $k = 5$ & 0$\cdot$21~(0$\cdot$01) & 49~(11) &
      1$\cdot$4~(0$\cdot$1) & 50~(11) & & 1$\cdot$40~(0$\cdot$02) &
      1$\cdot$9~(0$\cdot$7) & 0$\cdot$9~(0$\cdot$3) & 4$\cdot$2~(0$\cdot$8)\\
      $k = 10$ & 8~(2) & 76~(18) & 1$\cdot$4~(0$\cdot$1) & 85~(19) & & 
      12~(4) & 2$\cdot$4~(0$\cdot$8) & 1$\cdot$0~(0$\cdot$3) & 15~(4)\\
      $k = 25$ & 95~(38) & 117~(30) & 1$\cdot$4~(0$\cdot$1) & 214~(61)
      & & 86~(42) & 4~(1) & 1$\cdot$0~(0$\cdot$3) & 90~(43)\\
      $k = 50$ & 583~(313) & 348~(391) & 1$\cdot$5~(0$\cdot$1) &
      931~(535) & & 367~(222) & 62~(113) & 1$\cdot$0~(0$\cdot$3) &
      430~(262)
    \end{tabular}}
  \label{tab:CPUtimings}
  \begin{tabnote}
    Conditional simulations with $k=5$ do not use a Gibbs sampler.
  \end{tabnote}
\end{table}

Table~\ref{tab:CPUtimings} gives computational timings for conditional
simulations of max-stable processes on a $50 \times 50$ grid with a
varying number of conditioning locations. Due to the combinatorial
complexity of the partition set $\mathscr{P}_k$, the timings increase
rapidly with respect to the number of conditioning points $k$. It is
however reassuring that the algorithm is tractable when $k\in \{1,
\ldots, 50\}$; hence covering many practical situations and
applications.

\section{Application}
\label{sec:application}

\subsection{Extreme precipitations around Zurich}
\label{sec:extr-prec-around}

%% par("din") is 19.472222  5.175926 on my imac
\begin{figure}
  \centering
  \includegraphics[width=\textwidth]{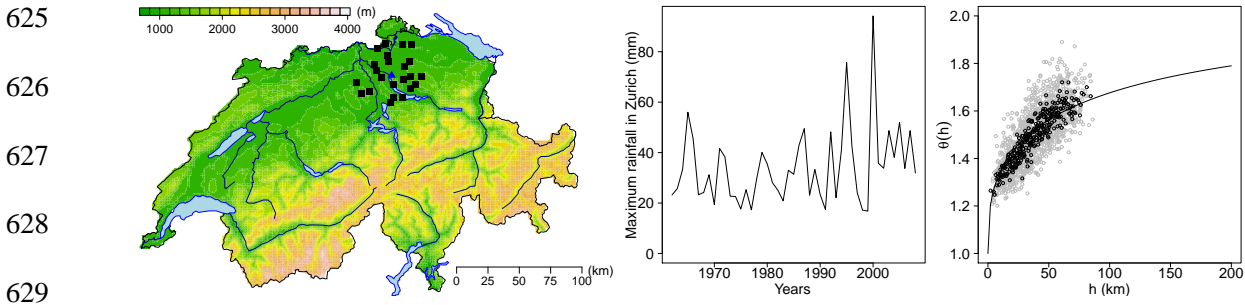}
  \caption{Left: Map of Switzerland showing the stations of the 24
    rainfall gauges used for the analysis, with an insert showing the
    altitude. The station marked with a triangle corresponds to
    Zurich. Middle: Summer maximum rainfall values for 1962--2008 at
    Zurich. Right: Comparison between the pairwise extremal
    coefficient estimates for the 51 original weather stations and the
    extremal coefficient function derived from a fitted Brown--Resnick
    processes having semi variogram $\gamma(h) = (h /
    \lambda)^\kappa$. The grey points are pairwise estimates; the black
    ones are binned estimates and the curve is the fitted extremal
    coefficient function.}
  \label{fig:prelimPlotRainfall}
\end{figure}

In this section we obtain conditional simulations of extreme
precipitation fields. The data considered here were previously
analyzed by \citet{Davison2011} who showed that Brown--Resnick
processes were one of the most competitive models among various
statistical models for spatial extremes.

The data are summer maximum rainfall for the years 1962--2008 at 51
weather stations in the Plateau region of Switzerland, provided by the
national meteorological service, MeteoSuisse. To ensure strong
dependence between the conditioning locations, we consider as
conditional locations the 24 weather stations that are at most 30km
apart from Zurich and set as the conditional values the rainfall
amounts recorded in the year 2000, the year of the largest
precipitation event ever recorded in Zurich between 1962--2008, see
Figure~\ref{fig:prelimPlotRainfall}. The largest and smallest
distances between the conditioning locations are around 55km and just
over 4km respectively.

A Brown--Resnick process having semi variogram $\gamma(h) = (h /
\lambda)^\kappa$ has to be fitted and the maximum pairwise likelihood
estimator introduced by \citet{Padoan2010} was used to simultaneously
fit the marginal parameters and the spatial dependence parameters
$\lambda$ and $\kappa$. In accordance with \citet{Davison2011}, the
marginal parameters were described by $\eta(x) = \beta_{0, \eta} +
\beta_{1, \eta} {\rm lon}(x) + \beta_{2, \eta} {\rm lat}(x)$,
$\sigma(x) = \beta_{0, \sigma} + \beta_{1, \sigma} {\rm lon}(x) +
\beta_{2, \sigma} {\rm lat}(x)$, $\xi(x) = \beta_{0, \xi}$, where
$\eta(x), \sigma(x), \xi(x)$ are the location, scale and shape
parameters of the generalized extreme value distribution and ${\rm
  lon}(x), {\rm lat}(x)$ the longitude and latitude of the stations at
which the data are observed. The maximum pairwise likelihood estimates
for $\lambda$ and $\kappa$ are 38 (14) and 0$\cdot$69 (0$\cdot$07) and
give a practical extremal range, i.e., the distance $h_+$ such that
$\theta(h_+) = \text{1$\cdot$7}$, of around 115km, see the right panel
of Figure~\ref{fig:prelimPlotRainfall}.

\begin{table}
  \tbl{Distribution of the partition size for the rainfall
    data estimated from a simulated Markov chain of length~15000}{%
    \begin{tabular}{lccccccc}
      Partition size & 1 & 2 & 3 & 4 & 5 & 6 & 7--24\\
      Empirical probabilities (\%) & 66$\cdot$2 & 28$\cdot$0 &
      4$\cdot$8 & 0$\cdot$5 & 0$\cdot$2 & 0$\cdot$2 & $<$0$\cdot$05
    \end{tabular}}\label{tab:partDistAppRainfall}
\end{table}

Table~\ref{tab:partDistAppRainfall} shows the distribution of the
partition size estimated from a Markov chain of length 15000. Around
65\% of the time the summer maxima observed at the 24 conditioning
locations were a consequence of a single extremal function, i.e., only
one storm event, and around 30\% of the time a consequence of two
different storms. Since the simulated Markov chain keeps a trace of
all the simulated partitions, we looked at the partitions of size two
and saw that around 65\% of the time, at least one of the four
up--north conditioning locations was impacted by one extremal function
while the remaining 20 locations were always influenced by another
one.\footnote{Mathieu: On a besoin de vérifier si cela colle aux
  données. J'ai pu avoir accès aux données journalières et les maxima
  estivaux pour l'année 2000 ont eu lieu les 5 (6), 11 (2) et 13 (8)
  juin, les 3 (1), 10 (2), 27 (1) et 29 (1) juillet et les 6 (1) et 27
  août (2). Les nombres entre parenthèses sont le nombre
  d'occurence. Il y a donc au total 9 événements différents mais 2
  sont vraiment très bien représentés: les 5 et 13 juin.}

\begin{figure}
  \centering
  \includegraphics[width=\textwidth]{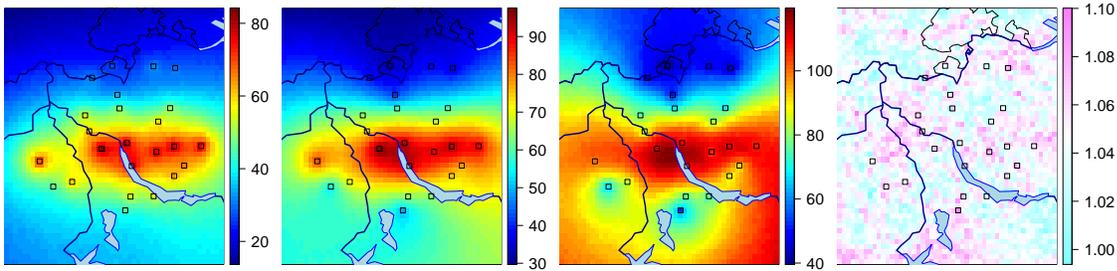}
  \caption{From left to right, maps on a $50 \times 50$ grid of the
    pointwise 0$\cdot$025, 0$\cdot$5 and 0$\cdot$975 sample quantiles
    for rainfall (mm) obtained from $10000$ conditional simulations of
    Brown--Resnick processes having semi variogram $\gamma(h) = (h /
    38)^{\text{0$\cdot$69}}$. The right most panel plots the ratio of
    the pointwise confidence intervals with and without taking into
    account the parameter estimate uncertainties. The squares show the
    conditional locations.}
  \label{fig:condSimAppRainfall}
\end{figure}

Figure~\ref{fig:condSimAppRainfall} plots the pointwise 0$\cdot$025,
0$\cdot$5 and 0$\cdot$975 sample quantiles obtained from $10000$
conditional simulations of our fitted Brown--Resnick process. The
conditional median provides a point estimate for the rainfall at an
ungauged location and the 0$\cdot$025 and 0$\cdot$975 conditional
quantiles a 95\% pointwise confidence interval.  As indicated by our
simulation study, see Figure~\ref{fig:condSim5condLoc2}, the shape
parameter $\kappa$ has a major impact on the regularity of paths and
on the width of the confidence interval. The value $\hat \kappa
\approx \text{0$\cdot$69}$ corresponds to very wiggly sample paths and
wider confidence intervals. To assess the impact of parameter
uncertainties on conditional simulations, the ratio of the width of
the confidence intervals with or without parameter uncertainty is
shown in the right panel of Figure~\ref{fig:condSimAppRainfall}. The
uncertainties were taken into account by sampling from the asymptotic
distribution of the maximum composite likelihood estimator and draw
one conditional simulation for each realization. These ratios show no
clear spatial pattern and the width of the confidence interval is
increased by an amount of at most 10\%.

\subsection{Extreme temperatures in Switzerland}
\label{sec:extr-temp-switz}

\begin{figure}
  \centering
  \includegraphics[width=\textwidth]{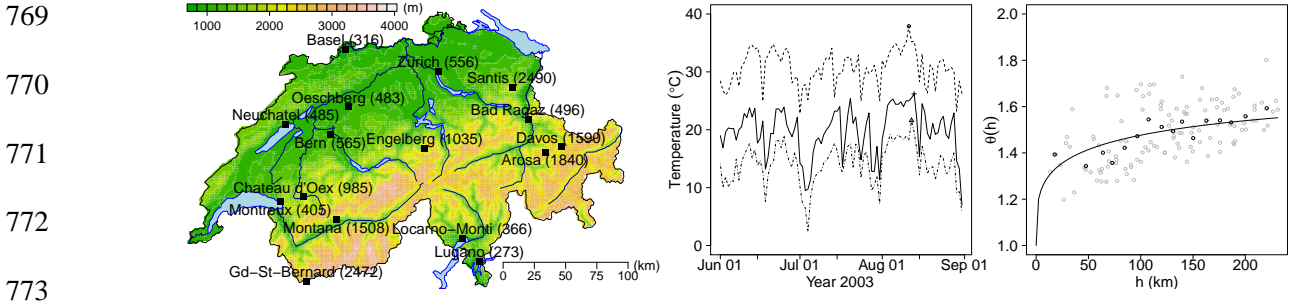}
  \caption{Left: Topographical map of Switzerland showing the sites
    and altitudes in metres above sea level of 16 weather stations for
    which annual maxima temperature data are available. Middle: Times
    series of the daily maxima temperatures at the 16 weather stations
    for year 2003. The '$\rm o$', '$+$' and '$\rm x$' symbols indicate
    the annual maxima that occurred the 11th, 12th and 13th of August
    respectively. Right: Comparison between the fitted extremal
    coefficient function from a Schlather process (solid red line) and
    the pairwise extremal coefficient estimates (gray circles). The
    black circles denote binned estimates with 16 bins.}
  \label{fig:studyRegionTemp}
\end{figure}

In this section we apply our results to get conditional simulations of
extreme temperature fields. The data considered here were previously
analyzed by \citet{Davison2011b} and consist in annual maximum
temperatures recorded at 16 sites in Switzerland during the period
1961--2005, see Figure~\ref{fig:studyRegionTemp}.

Following the work of \citet{Davison2011b}, we fit a Schlather process
with an isotropic powered exponential correlation function and trend
surfaces $\eta(x) = \beta_{0, \eta} + \beta_{1, \eta} {\rm alt}(x)$,
$\sigma(x) = \beta_{0, \sigma}$, $\xi(x) = \beta_{0,\xi} + \beta_{1,
  \xi} {\rm alt}(x)$, where ${\rm alt}(x)$ denotes the altitude above
mean sea level in kilometres and $\{\eta(x), \sigma(x), \xi(x)\}$ are
the location, scale and shape parameters of the generalized extreme
value distribution at location $x$. The spatial dependence parameter
estimates are $\hat \lambda = 260~(149)$ and $\hat \kappa =
0.52~(0.12)$ and the corresponding fitted extremal coefficient
function, similar to some extent to our test case $\rho_3$ in
Section~\ref{sec:simulation-study}, is shown in the right panel of
Figure~\ref{fig:studyRegionTemp}.

\begin{table}
  \tbl{Distribution of the partition size for the temperature
    data estimated from a Markov chain of length 10000}{%10 Markov
                                %chains of length~1000}{%
    \begin{tabular}{lccccc}
      Partition size & 1 & 2 & 3 & 4 & 5--16\\
      Empirical probabilities (\%) & 2$\cdot$47 & 21$\cdot$55 &
      64$\cdot$63 & 10$\cdot$74 & 0$\cdot$61\\
    \end{tabular}}\label{tab:partDistAppTemp}
\end{table}

In year 2003, western Europe was hit by a severe heat wave believed to
be the hottest one ever recorded since at most 1540 (``2003 European
heat wave'', Wikipedia: The Free Encyclopedia). Switzerland was
largely impacted by this severe extreme event since the nation wide
record temperature of 41$\cdot$5$^\circ{\rm C}$ was recorded that year
in Grono, Graubunden, near Lugano. Consequently for our analysis we
use as conditional event the maxima temperatures observed in summer
2003, see Figure~\ref{fig:studyRegionTemp}. Based on the fitted
Schlather model, we simulate a Markov chain of effective length
10000% 10 parallel Markov chains of effective
% length 1000,
with a burn-in period of length 500 and a thinning lag of
100 iterations. The distribution of the partition size estimated from
these Markov chains is shown in Table~\ref{tab:partDistAppTemp}.  We
can see that around 90\% of the time the conditional realizations were
a consequence of at most three extremal functions. Since our original
observations were not summer maxima but maximum daily values, a close
inspection of the times series in year 2003 reveals that the hottest
temperatures occurred between the 11th and 13th of August, see
Figure~\ref{fig:studyRegionTemp}, and, to some extent, corroborates
the distribution of Table~\ref{tab:partDistAppTemp}.

%% par("din") is 18.759259  2.925926 on my iMac
\begin{figure}
  \centering
  \includegraphics[width=\textwidth]{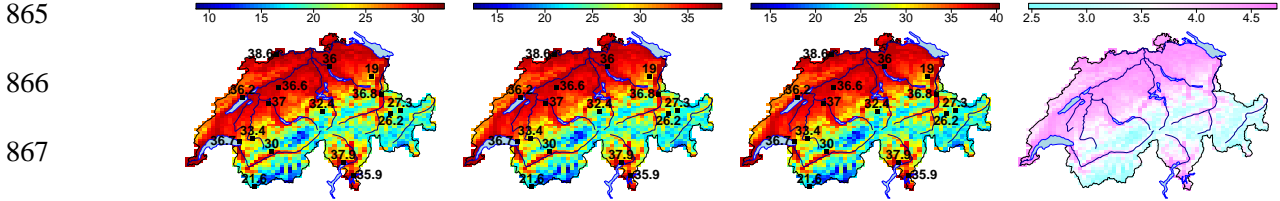}
  \caption{From left to right: Maps on a $64 \times 64$ grid of the
    pointwise 0$\cdot$025, 0$\cdot$5 and 0$\cdot$975 sample quantiles
    for temperature $(^\circ{\rm C})$ obtained from 10000 conditional
    simulations of the fitted Schlather process. The squares show the
    conditional locations and the associated conditional values. The
    right most panel shows temperature anomalies, i.e., the difference
    between the pointwise conditional and unconditional medians.}
  \label{fig:condSimAppTemp}
\end{figure}

Figure~\ref{fig:condSimAppTemp} shows the 0$\cdot$025, 0$\cdot$5 and
0$\cdot$975 pointwise sample quantiles obtained from 10000 conditional
simulations on a $64 \times 64$ grid. As expected, we can see that the
largest temperatures occurred in the plateau region of Switzerland
while temperatures were appreciably cooler in the Alps. The right
panel of Figure~\ref{fig:condSimAppTemp} shows the difference between
the pointwise conditional medians and the pointwise unconditional
medians estimated from the fitted trend surfaces. The differences
range between 2$\cdot$5$^\circ{\rm C}$ and 4$\cdot$75$^\circ{\rm C}$
and the largest deviations occur in the plateau region of Switzerland.

\section*{Acknowledgements}
\label{sec:acknowledgments}

M. Ribatet was partly funded by the MIRACCLE-GICC and McSim ANR
projects. The authors thank MeteoSuisse and Dr. S.~A.~Padoan for
providing the precipitation data and Prof. A.~C.~Davison and
Dr. M.~M.~Gholam-Rezaee for providing the temperature data set.

% \section*{Supplementary material}
% \label{sec:suppl-mater}

% Supplementary material available at \emph{Biometrika} online includes
% \textbf{to be completed later!!!}

\appendix
\section*{Appendix}

\subsection*{The Brown--Resnick model}
For all $x \in \mathcal{X}^k$ and Borel set $A \subset \mathbb{R}^k$
\begin{equation*}
  \Lambda_x(A) = \int_0^\infty \Pr\left[\zeta \exp\{W(x)-\gamma(x)\}
    \in A\right] \zeta^{-2} \mbox{d$\zeta$} = \int_0^\infty
  \int_{\mathbb{R}^k} 1_{\{\zeta \exp\{y-\gamma(x)\} \in A \}}
  f_x(y) \mbox{d$y$} \zeta^{-2} \mbox{d$\zeta$},
\end{equation*}
where $f_x$ denotes the density of the random vector $W(x)$, i.e., a
centered Gaussian random vector with covariance matrix $\Sigma_x$ and
variance $2 \gamma(x)$. The change of variables $z=\zeta
\exp\{y-\gamma(x)\}$ and $r=\log \zeta$ yields
\begin{equation*}
  \Lambda_x(A) =  \int_{-\infty}^\infty \int_A f_x\{\log z - r + \gamma(x)
  \} \prod_{i=1}^k z_i^{-1}\mbox{d$z$} e^{-r} \mbox{d$r$} = \int_{A}
  \lambda_x(z) \mbox{d$z$}
\end{equation*}
with 
\begin{equation*}
  \lambda_x(z) =  \prod_{i=1}^k z_i^{-1} \int_{-\infty}^\infty  f_x\{\log z
  - r + \gamma(x)\} e^{-r} \mbox{d$r$}.
\end{equation*}
Since
\begin{equation*}
f_x\{\log z - r + \gamma(x)\} e^{-r}= (2 \pi)^{-k/2} |\Sigma_x|^{-1/2}
\exp\left\{-\frac{1}{2}P(r)\right\},
\end{equation*}
with
\begin{equation*}
  P(r) = r^2 1_k^T \Sigma_x^{-1} 1_k - 2 r \left[1_k^T\Sigma_x^{-1}\{\log 
  z + \gamma(x)\} - 1\right] + \{\log z + \gamma(x)\}^T \Sigma_x^{-1}
\{\log z + \gamma(x)\},
\end{equation*}
standard computations for Gaussian integrals give
\begin{equation*}
  \lambda_x(z) = C_x \exp\left(-\frac{1}{2} \log {z}^T Q_x \log 
    z + L_x \log z \right) \prod_{i=1}^k z_i^{-1}.
\end{equation*}
The conditional intensity function is
\begin{equation*}
  \lambda_{s \mid x, z}(u) %&=\frac{\lambda_{(s,x)}(u,z)}{\lambda_{x}(z)}\\  
  = \frac{C_{(s,x)}}{C_x} \exp\left\{-\frac{1}{2} \log {(u,z)}^T
    Q_{(s,x)} \log  (u,z) + L_{(s,x)} \log (u,z)+\frac{1}{2} \log
    {z}^T Q_x \log z - L_x \log z \right\} \prod_{i=1}^m u_i^{-1},
\end{equation*}
and since $\log(u,z)=J_{m,k}\log u+\tilde J_{m,k}\log z$, it is not
difficult to show that
\begin{equation*}
  \lambda_{s \mid x, z}({u}) = \frac{C_{(s,x)}}{C_x}
  \exp\left\{-\frac{1}{2} (\log u - \mu_{s \mid x, z})^T \Sigma_{s
      \mid x}^{-1} (\log u - \mu_{s \mid x, z}) \right\} \prod_{i=1}^m
  u_i^{-1}.
\end{equation*}
Finally, the relation $C_{(s,x)}/C_x= (2\pi)^{-m/2} |\Sigma_{s \mid
  x}|^{-1/2}$ is a simple consequence of the normalization
$\int\lambda_{s \mid x, z}({u})\mbox{d$u$}=1$.

\subsection*{The Schlather model}

For all $x \in \mathcal{X}^k$ and Borel set $A \subset \mathbb{R}^k$
\begin{equation*}
  \Lambda_x(A) = \int_0^\infty \Pr[\sqrt{2\pi} \zeta \varepsilon(x)
  \in A] \zeta^{-2} \mbox{d$\zeta$} = \int_0^\infty
  \int_{\mathbb{R}^k} 1_{\left\{\sqrt{2\pi} \zeta y \in A \right\}}
  f_{\mathbf{x}}(y) \mbox{d$y$} \zeta^{-2} \mbox{d$\zeta$},
\end{equation*}
where $f_x$ denotes the density of the random vector $\varepsilon(x)$,
i.e., a centered Gaussian random vector with covariance matrix
$\Sigma_x$. The change of variable $z = \sqrt{2\pi} \zeta y$ gives
\begin{align*}
  \Lambda_{x}(A) &= (2 \pi)^{-k/2} \int_0^\infty \int_A
  f_x\left(\frac{z}{\sqrt{2 \pi} \zeta} \right) \zeta^{-(k+2)}
  \mbox{d$z$d$\zeta$}\\
  &= (2 \pi)^{-k} |\Sigma_x|^{-1/2} \int_0^\infty \int_A \exp\left(-
    \frac{1}{4 \pi \zeta^2} z^T \Sigma_x^{-1} z \right) \zeta^{-(k+2)}
  \mbox{d$z$d$\zeta$}\\
  % &= (2 \pi)^{-k} |\Sigma_x|^{-1/2}  \int_A \int_0^\infty \exp\left(-
  %   \frac{\zeta^2}{4 \pi} z^T \Sigma_x^{-1} z \right) \zeta^k
  % \mbox{d$\zeta$d$z$}\\
  &= (2 \pi)^{-k} |\Sigma_x|^{-1/2} \int_A \frac{2 \pi}{z^T
    \Sigma_x^{-1} z} \mathbb{E}[X^{k-1}] \mbox{d$z$}, \qquad 
  X \sim \text{Weibull}\left(\sqrt{\frac{4 \pi}{z^T \Sigma_x^{-1} z}},
    2 \right)\\ 
  &= (2 \pi)^{-k} |\Sigma_x|^{-1/2} \int_A \frac{2 \pi}{z^T
    \Sigma_x^{-1} z} \left(\frac{4 \pi}{z^T \Sigma_x^{-1} z}
  \right)^{(k-1)/2} \Gamma\left(\frac{k+1}{2} \right) \mbox{d$z$}\\
  &= \int_A \lambda_\mathbf{x}(z) \mbox{d$z$},
\end{align*}
where $\lambda_x(z) = \pi^{-(k - 1)/2} |\Sigma_x|^{-1/2}
a_x(z)^{-(k+1)/2} \Gamma\left\{(k+1)/2 \right\}$ and $a_x(z) = z^T
\Sigma_x^{-1} z$.

For all $u \in \mathbb{R}^m$ the conditional intensity function is
\begin{equation*}
  \label{eq:conditionalDensity}
  \lambda_{s \mid x, z}(u) = \pi^{-m/2} \frac{|\Sigma_{(s,
      x)}|^{-1/2}}{|\Sigma_x|^{-1/2}} \left\{\frac{a_{(s, x)}(u,
      z)}{a_x(z)} \right\}^{-(m+k+1)/2} a_x(z)^{-m/2}
  \frac{\Gamma\left(\frac{m+k+1}{2}\right)}{\Gamma\left(\frac{k+1}{2}\right)}.
\end{equation*}

We start by focusing on the ratio $a_{(s, x)}(u, z) / a_x(z)$. Since
\begin{equation*}
  \begin{bmatrix}
    \Sigma_s & \Sigma_{s:x}\\
    \Sigma_{x:s} & \Sigma_x
  \end{bmatrix}^{-1}
  =
  \begin{bmatrix}
    (\Sigma_s - \Sigma_{s:x} \Sigma_x^{-1} \Sigma_{x:s})^{-1} & -
    (\Sigma_s - \Sigma_{s:x} \Sigma_x^{-1} \Sigma_{x:s})^{-1}
    \Sigma_{s:x} \Sigma_x^{-1}\\
    - \Sigma_x^{-1} \Sigma_{x:s} (\Sigma_s - \Sigma_{s:x}
    \Sigma_x^{-1} \Sigma_{x:s})^{-1} & \Sigma_x^{-1} + \Sigma_x^{-1}
    \Sigma_{x:s} (\Sigma_s - \Sigma_{s:x} \Sigma_x^{-1}
    \Sigma_{x:s})^{-1} \Sigma_{s:x} \Sigma_x^{-1}
  \end{bmatrix},
\end{equation*}
straightforward algebra shows that
\begin{equation*}
  \frac{a_{(s, x)}(u, z)}{a_x(z)} = 1 +\frac{(u - \mu)^T \tilde
    \Sigma^{-1} (u - \mu)}{k+1}, \qquad  \mu = \Sigma_{s:x}
  \Sigma_x^{-1} z, \qquad \tilde \Sigma = \frac{a_x(z)}{k+1}
  \left(\Sigma_s - \Sigma_{s:x} \Sigma_x^{-1} \Sigma_{x:s} \right).
\end{equation*}

We now try to simplify the ratio $|\Sigma_{(s, x)}|
/|\Sigma_x|$. Using the fact that
\begin{equation*}
  \Sigma_{(s, x)} =
  \begin{bmatrix}
    \Sigma_s & \Sigma_{s:x}\\
    \Sigma_{x:s} & \Sigma_x
  \end{bmatrix}
  =
  \begin{bmatrix}
    \mbox{Id}_m & \Sigma_{s:x}\\
    0_{k,m} &  \Sigma_x
  \end{bmatrix}
  \begin{bmatrix}
    \Sigma_s - \Sigma_{s:x} \Sigma_x^{-1} \Sigma_{x:s} & 0_{m,k}\\
    \Sigma_x^{-1} \Sigma_{x:s} & \mbox{Id}_k
  \end{bmatrix},
\end{equation*}
combined with some more algebra yields
\begin{equation*}
  \frac{|\Sigma_{(s, x)}|}{|\Sigma_x|} = |\Sigma_s - \Sigma_{s:x}
  \Sigma_x^{-1} \Sigma_{x:s}| = \left\{\frac{k+1}{a_x(z)}\right\}^m
  |\tilde \Sigma|.
\end{equation*}

Using the two previous results it is easily found that
\begin{equation*}
  \lambda_{s \mid x, z}(u) = \pi^{-m/2} (k+1)^{-m/2} |\tilde
  \Sigma|^{-1/2} \left\{1 + \frac{(u - \mu)^T \tilde \Sigma^{-1} (u -
      \mu)}{k+1} \right\}^{-(m+k+1)/2}
  \frac{\Gamma\left(\frac{m+k+1}{2}
    \right)}{\Gamma\left(\frac{k+1}{2}\right)},
\end{equation*}
which corresponds to the density of a multivariate Student
distribution with the expected parameters.

% \bibliographystyle{biometrika}
% \bibliography{/Users/Mathieu/Documents/Biblio/biblio_ribatet,biblio}
\end{document}